\documentclass{PoS}

\usepackage[utf8]{inputenc}

\newcommand{\be}{\begin{equation}}
\newcommand{\ee}{\end{equation}}
\newcommand{\bea}{\begin{eqnarray}}
\newcommand{\eea}{\end{eqnarray}}
\newcommand{\bean}{\begin{eqnarray*}}
\newcommand{\eean}{\end{eqnarray*}}
\newcommand{\nn}{\nonumber}

\newcommand{\C}{\mathbb{C}}

\newcommand{\xv}{{\mathbf x}}

\title{On complex Langevin dynamics and zeroes of the measure II: Fermionic determinant}

\ShortTitle{On complex Langevin dynamics and zeroes of the measure II: Fermionic determinant}

\author{Gert Aarts\\
        Department of Physics, College of Science, Swansea University, Swansea, United Kingdom  \\
        E-mail: \email{g.aarts@swan.ac.uk}}

\author{Erhard Seiler\\
 Max-Planck-Institut for Physics (Werner-Heisenberg-Institut) Munich, Germany\\
        E-mail: \email{ehs@mppmu.mpg.de}}

\author{\speaker{D\'enes Sexty} \thanks{We would like to thank Felipe Attanasio, Zolt\'an Fodor, Benjamin J\"ager, S\'andor Katz and Csaba T\"or\"ok 
for collaboration on related topics. }\\
     Department of Physics, University of Wuppertal, Wuppertal, Germany \\
     Institute for Theoretical Physics, E\"otv\"os University, Budapest, Hungary\\
        E-mail: \email{sexty@uni-wuppertal.de}}

\author{Ion-Olimpiu Stamatescu\\
        Institute for Theoretical Physics, University of Heidelberg and FEST, Heidelberg, Germany\\
        E-mail: \email{I.O.Stamatescu@thphys.uni-heidelberg.de}}

\abstract{Lattice QCD at non-vanishing chemical potential is studied using the complex
 Langevin equation (CLE).
One of the conditions for the correctness of the results of the CLE
is that the zeroes of the measure coming from the fermionic determinant are outside of the distribution of 
the configurations, or at least in a
region where support for the distribution is very much suppressed.
We investigate this issue for Heavy Dense QCD (HDQCD) 
and 
full QCD at high temperatures. 
In HDQCD it is found that the configurations move closest to the zeroes of the measure 
around the critical chemical potential of the onset transition, where the sign problem is diminished, 
but results remain largely unaffected.
In full QCD at high temperatures the investigation of the spectrum of the 
Dirac operator 
yields a similar observation: the results are unaffected by the issue of the poles.
}

\FullConference{34th annual International Symposium on Lattice Field Theory\\
		24-30 July 2016\\
		University of Southampton, UK}

\begin{document}

\section{Introduction}

At nonzero baryon chemical 
potential, QCD suffers from the sign problem which invalidates naive
importance sampling simulations \cite{deForcrand:2010ys}.
The Complex Langevin Equation (CLE) was proposed to circumvent the sign problem 
by complexifying the field manifold of the theory \cite{Parisi:1984cs}.
In the past few years the method has been the focus of renewed attention \cite{Aarts:2008rr,Aarts:2008wh,Mollgaard:2013qra,Fodor:2015doa} (see further references in \cite{deForcrand:2010ys,Sexty:2014dxa}). In particular with the introduction of gauge cooling 
\cite{gaugecooling} it has become possible to simulate gauge theories
as well, such as heavy dense QCD (HDQCD) \cite{gaugecooling,Aarts:2016qrv} and full QCD using light quark flavors \cite{Sexty:2013ica,Aarts:2014bwa}.

In fermionic theories the action includes a non-holomorphic term (the logarithm 
of the fermionic determinant) which might cause problems 
\cite{Mollgaard:2013qra}.
Here we investigate this issue for lattice models. The presentation is based on
\cite{polepaperinprep} and is accompanied by \cite{gert2016proc} where the 
theoretical background 
is discussed in more detail and toy models are investigated.

\section{CLE and poles}

Consider a theory with measure $\rho(x)$ where $x$ stands for all the degrees of 
freedom of the theory. Typically one has $ \rho(x) = e^{-S(x)} \det M(x) $ with 
the action $S(x)$ and the determinant of the Dirac matrix $M(x)$. 
The Langevin equation is then written as 
\bea
 { \partial x \over  \partial\tau} = K(x) + \eta(\tau) \ \ \ \qquad \qquad 
 K(x) = \rho'(x) /\rho(x)  
\eea
with the Langevin time $\tau$, delta correlated Gaussian noise $\eta (\tau)$,
and the drift term $K(z)$. (For curved manifolds such as the $SU(3)$ group space this
equation has to be appropriately modified, see e.g. \cite{Aarts:2008rr}.)
For complex measures the manifold of $x$ is complexified, for example in QCD
SU(3) is extended into SL(3,$\C$).

Note that $K(z)$ has a singularity where the measure vanishes. 
This in turn invalidates the proof of correctness, which requires fast decay 
of the distributions of the fields on the complexified manifold as well 
as holomorphic drift terms (and thus also action) on the whole 
complex plane \cite{Aarts:2009uq}.
In \cite{polepaperinprep,gert2016proc} it is shown that the proof still holds 
as long as the zeroes of the measure are outside of the support of the 
distribution of $x$ on the complexified manifold. In the case where the 
zeroes are inside the distribution one typically observes a 'bottleneck' phenomenon
where the zeroes connect two disjoint regions of the configuration space visited 
during the Langevin process.

\section{Heavy Dense QCD}

HDQCD is defined as the double limit $ \kappa \rightarrow 0 $ and 
$ \mu \rightarrow \infty $ while keeping $ \kappa e^\mu $ fixed \cite{Bender:1992gn}. 
For symmetry
we keep terms proportional to $ \kappa e^{-\mu}$. The fermionic
determinant then simplifies to 
\be \label{eq:hddet}
 \det M = \prod_\xv
                \det\left(1 + h e^{\mu/T}  \mathcal{P}_\xv \right)^{2N_f}
                \det\left( 1 + h e^{-\mu/T}  \mathcal{P}^{-1}_\xv \right)^{2N_f}
\end{equation}
for $N_f$ fermionic flavors, using the Polyakov loop $ \mathcal{P}_\xv $ and  $h=(2\kappa)^{N_{\tau}}$.


This theory has been studied extensively for the justification 
of the CLE approach by comparing to reweighting 
\cite{gaugecooling}, and the phase diagram has been 
mapped out in \cite{Aarts:2016qrv}.

\begin{figure}[t]
\begin{center}
\includegraphics[width=0.49\columnwidth]{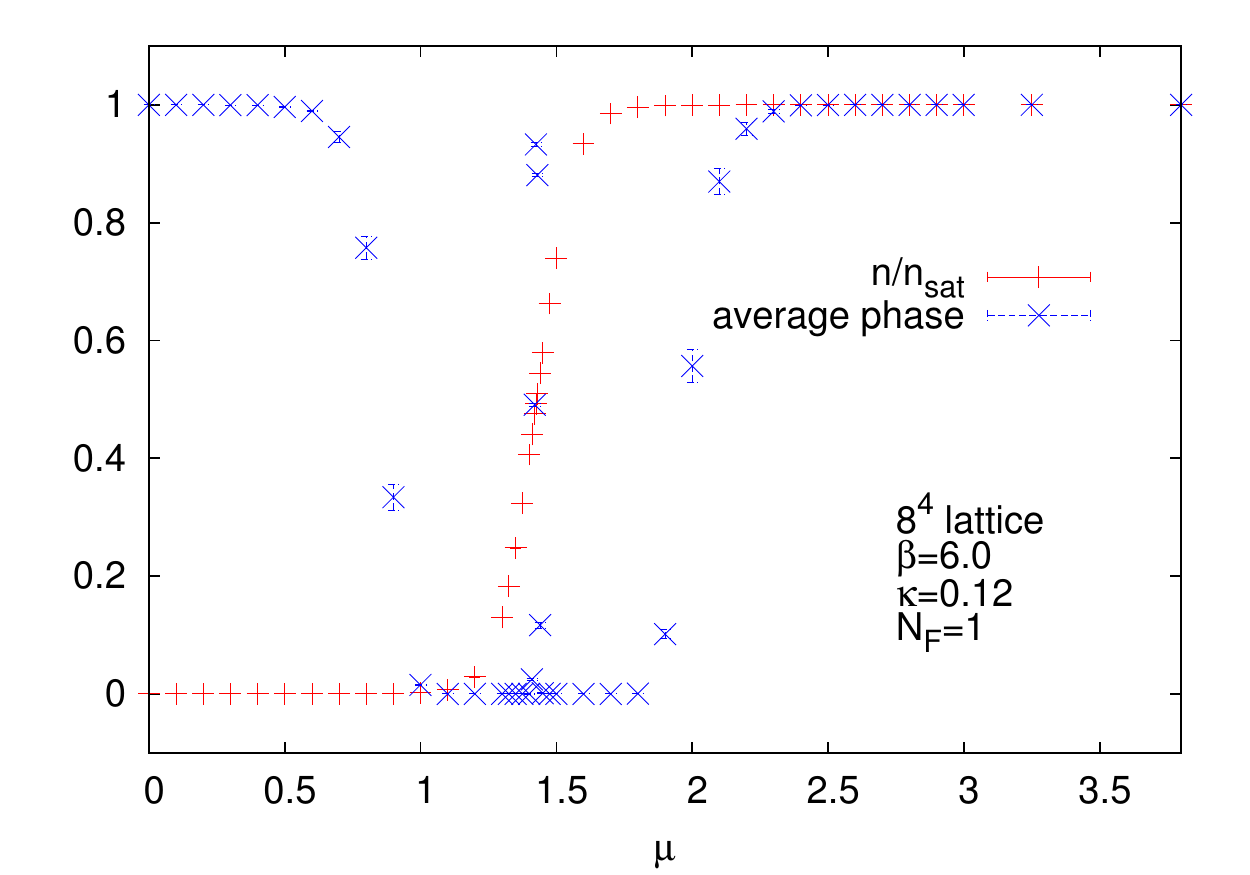}
\includegraphics[width=0.49\columnwidth]{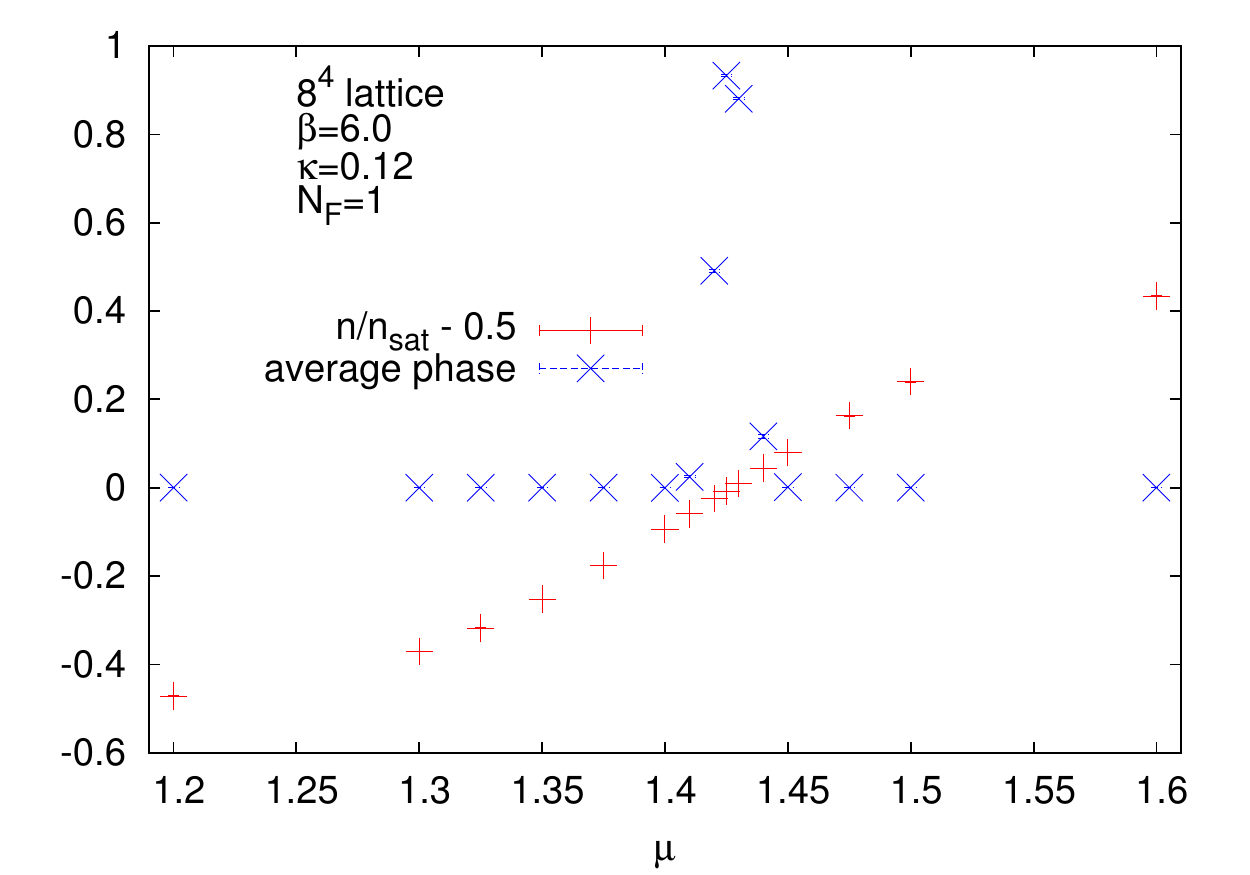}
\caption{Left: fermionic density in units of the saturation density, $n/n_{\rm sat}$, 
and average phase factor, 
see Eq.\ (\protect\ref{signaver}), as a function of the chemical potential. 
Right: a blow-up around onset. 
}
\label{HDsigndens}
\end{center}
\end{figure}

The quark density in HDQCD at low temperatures shows a Silver-Blaze behavior 
as expected, with an onset transition at the critical chemical potential 
\be
 \mu_c =-\ln (2\kappa),
\ee
and eventually saturating at the value $n_{\rm sat}=6N_f$. At higher 
temperatures this behavior is somewhat smoothened, with roughly constant
critical chemical potential up to moderate temperatures \cite{Aarts:2016qrv}.
In Fig.\ \ref{HDsigndens} the quark density is shown in units of saturation density, for $\kappa=0.12, N_f=1, \beta=6.0 $ on a $8^4$ lattice, with 
$\mu_c=-\ln(2 \kappa) \approx 1.427$. 
Using $ [\det M(\mu)]^* = \det M(-\mu^*)$ we also measure the average phase 
factor in the full theory using the analytic observable
\be
 \label{signaver}
 \langle e^{2i\varphi} \rangle = \left\langle \frac{\det M(\mu)}{\det M(-\mu)} \right\rangle,
\ee
see also Fig.\ \ref{HDsigndens}. 
One notes a severe sign problem except in the initial 
zero density phase as well as in saturation and a peak in the middle. 
At the critical chemical potential, the fermions are at `half-filling', i.e.\
half of the available fermionic states are filled, and the theory
becomes particle-hole symmetric \cite{Rindlisbacher:2015pea}.
Exactly at $\mu_c$, the sign problem becomes very mild, as the first 
factor in Eq.\ (\ref{eq:hddet}) becomes real.
The sign problem due to the second factor is mild, since $ (2\kappa e^{-\mu})^{2N_\tau}\ll 1$.

\begin{figure}[t]
\begin{center}
\includegraphics[width=0.49\columnwidth]{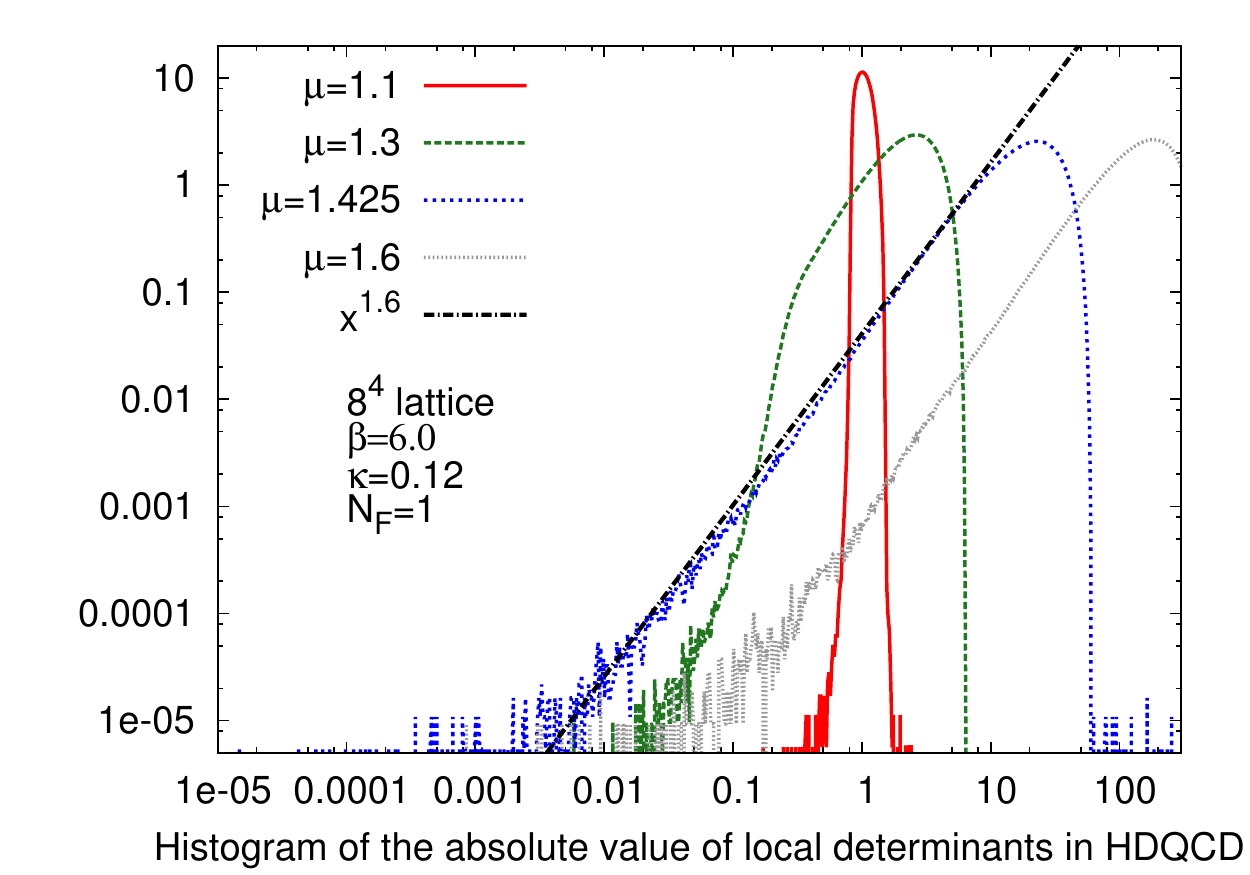}
\includegraphics[width=0.49\columnwidth]{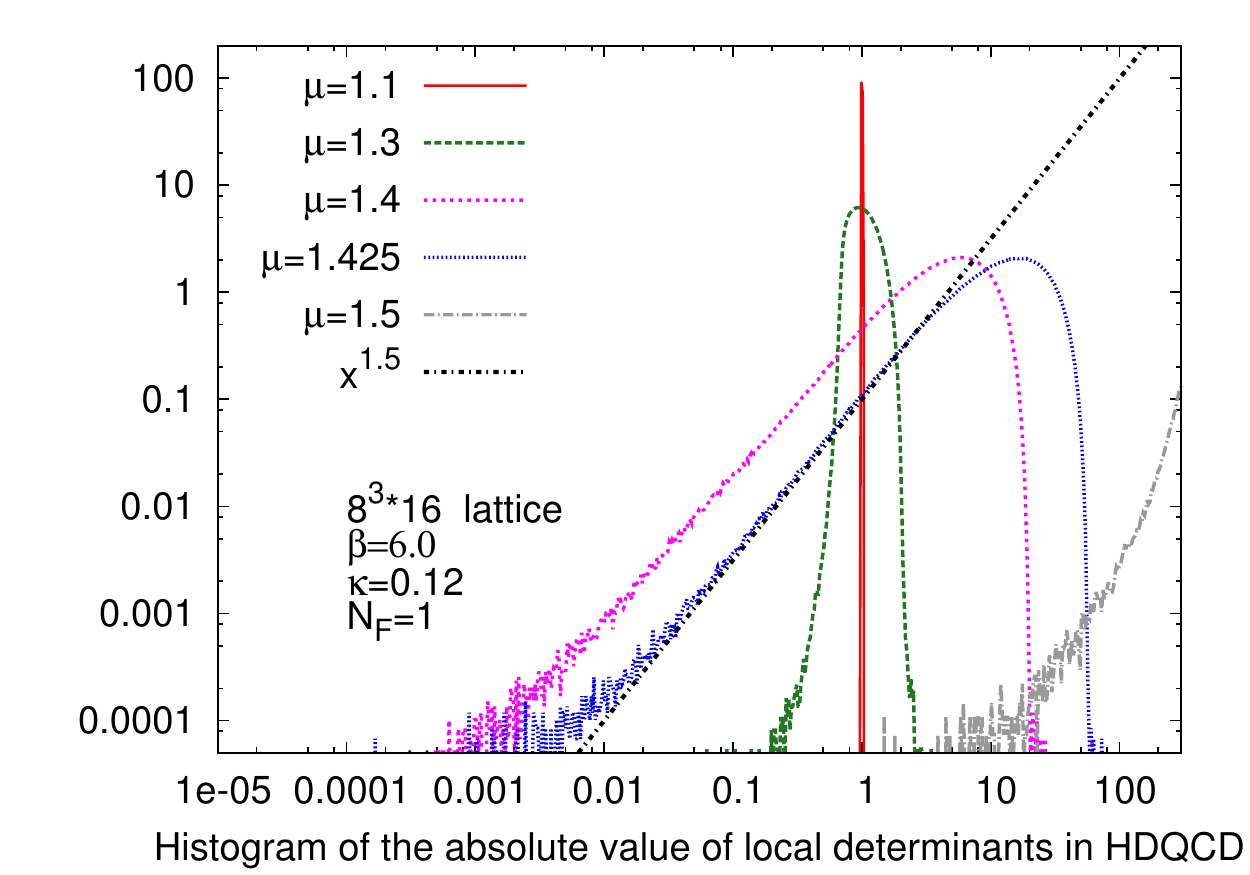}
\caption{The histogram of the absolute value of the local determinant 
factors (\protect\ref{eq:detloc}). 
}
\label{rhist}
\end{center}
\end{figure}

\begin{figure}[t]
\begin{center}
\includegraphics[width=0.49\columnwidth]{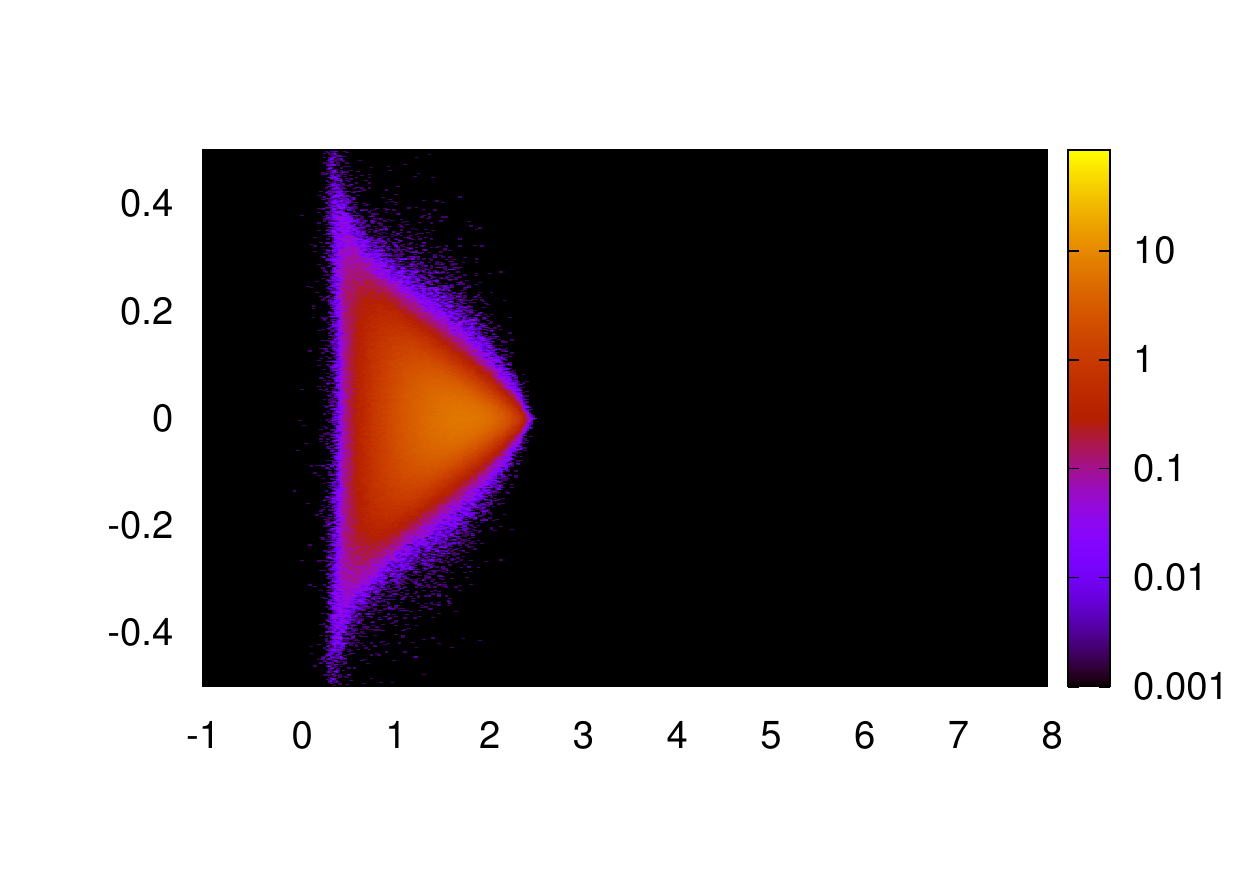}
\includegraphics[width=0.49\columnwidth]{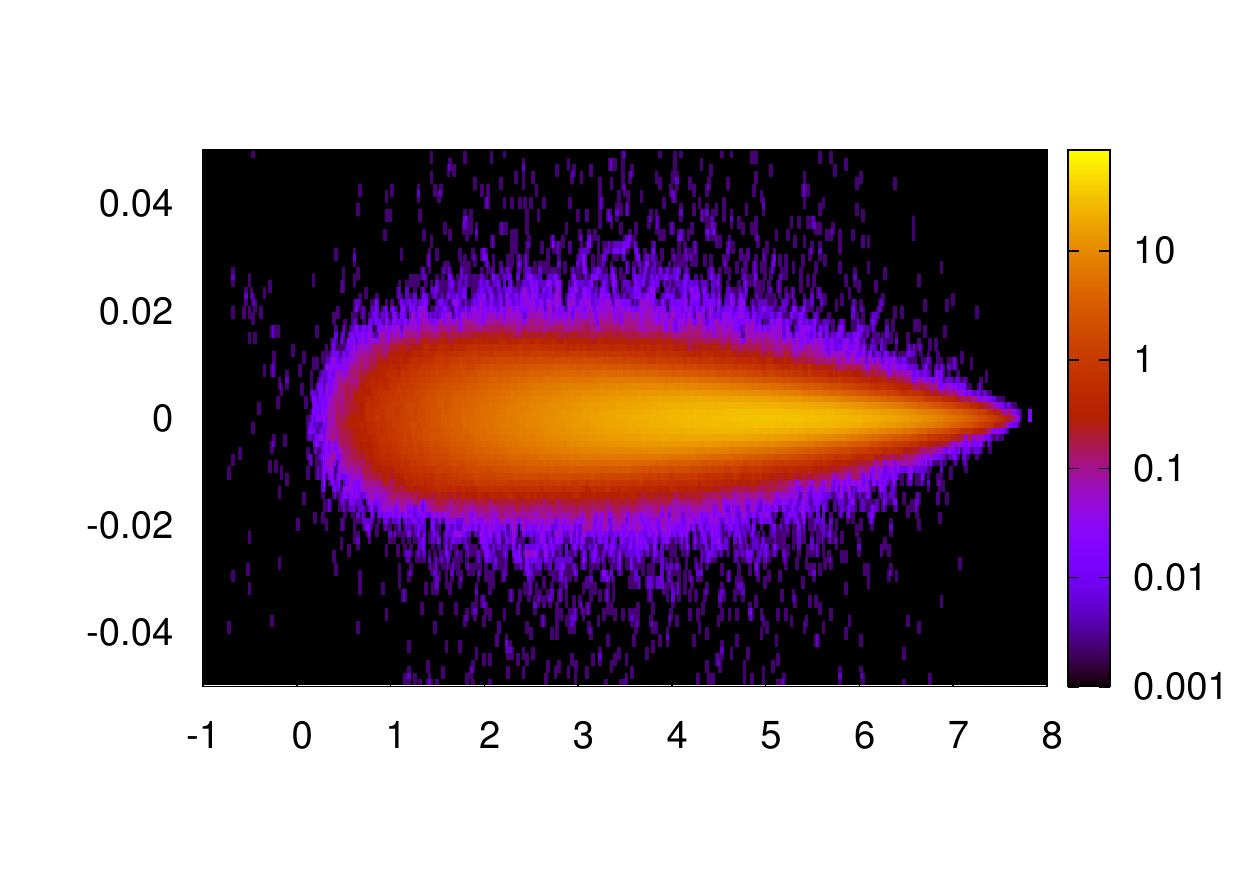}
\caption{The histogram of the absolute value of the local determinant 
factors (\protect\ref{eq:detloc}). 
}
\label{hd2dhist}
\end{center}
\end{figure}

To investigate the density of configurations around the zeroes of the 
measure we investigated the spatially local factors in the 
fermionic determinant, namely 
\bea
\det 
M_\xv 
&=&
\det\left(1 + h e^{\mu/T}  \mathcal{P}_\xv \right) \det\left( 1 + h e^{-\mu/T} 
 \mathcal{P}^{-1}_\xv \right)
\nn\\
    &=& \left(1+3z P_\xv+3z^2P_\xv^{-1}+z^3\right) \left(1+3\bar z P_\xv^{-1} +3\bar z^2P_\xv+\bar z^3\right),
\label{eq:detloc}
\eea
with $z=he^{\mu/T}, \bar z=he^{-\mu/T}$. We find that the determinants largely 
avoid the zeroes of the measure, except close to the critical 
chemical potential. In Fig.~\ref{rhist} we show the histogram of the 
absolute values of the determinants for two different lattice sizes.
Far from $\mu_c$ one observes that the zero is far from the distribution. Around the 
critical chemical potential one observes a power law dependence of 
the histogram on 
the distance from the origin $ P \sim |\det M|^\alpha $ with 
an exponent $ \alpha=1.5-1.6 $.
It is interesting to note that in the histogram of $ \Phi$, the phases of 
the local determinants also show a power law behavior around $\mu_c$ with 
the dependence
$ P \sim 1/ \Phi^3 $. The phase of the total determinant is the sum of 
$ 2 N_f N_s^3$ phases, and will therefore show large fluctuations as the volume 
increases. In Fig.~\ref{hd2dhist} we show the distribution 
of the local determinant factors on the complex plane.
Comparing with results of various toy models,
we do not observe disconnected regions separated 
by the singular point which would signal incorrect results.
In particular in a one plaquette model with the same determinant 
factor in the measure, the bad behavior is 
signaled by 'whiskers' in the histograms \cite{polepaperinprep,gert2016proc}, which 
appear here only 
barely visible, if at all.

To summarize, in HDQCD we see that far from the critical chemical potential,
where the sign problem is severe, we see no issue from the poles. 
Curiously, near $\mu_c$ where the sign problem is mild, the distribution
of the configurations gets closest to the zeroes, but 
for the parameters used here we see no sign of problematic disconnected 
regions in the observables suggesting that the effect of the non-analyticity
is very small.

\section{Full QCD}

Finally we investigate the issue of poles for full QCD in the staggered 
formulation using $N_f=4$, i.e. not using the rooting procedure to 
further reduce the number of flavors described by the Dirac operator.
Several observations show that at high temperatures 
(above the deconfinement transition) the CLE is unaffected by the presence 
of zeroes in the measure: comparisons with systematic hopping expansions using 
a holomorphic formulation \cite{Aarts:2014bwa},
comparisons with reweighting \cite{Fodor:2015doa}, and observations 
of the spectra of the complexified Dirac operator 
\cite{Sexty:2014dxa,polepaperinprep}.

\begin{figure}[t]
\begin{center}
\includegraphics[width=0.49\columnwidth]{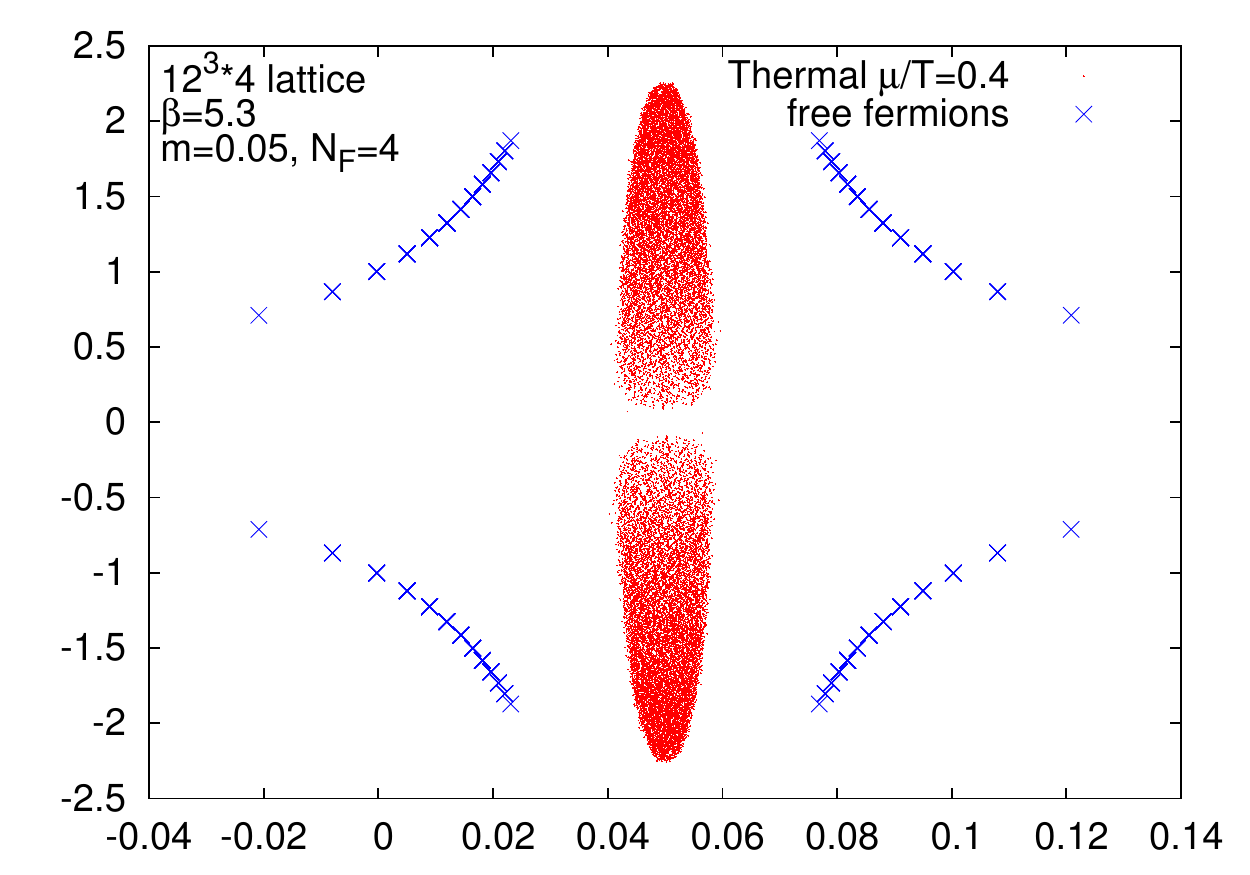}
\includegraphics[width=0.49\columnwidth]{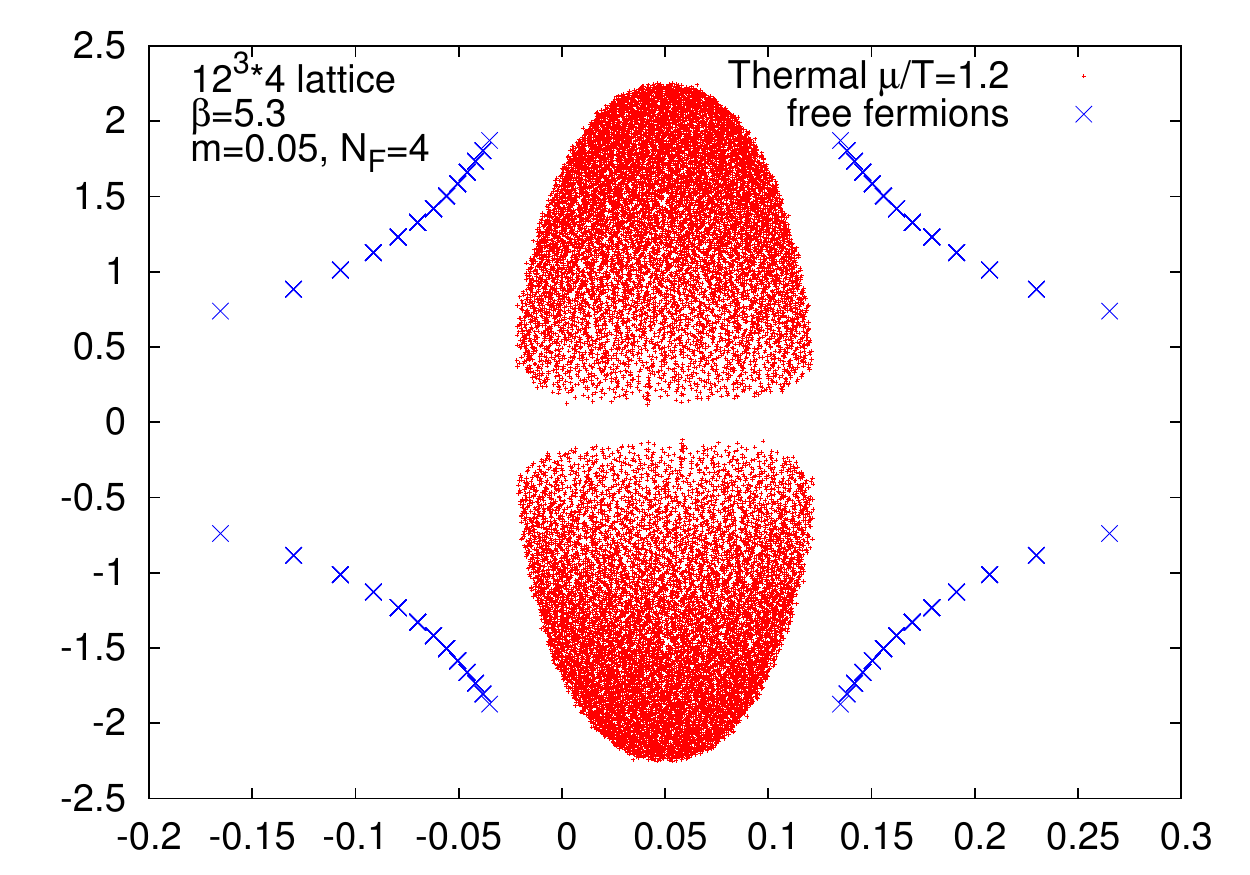}
\includegraphics[width=0.49\columnwidth]{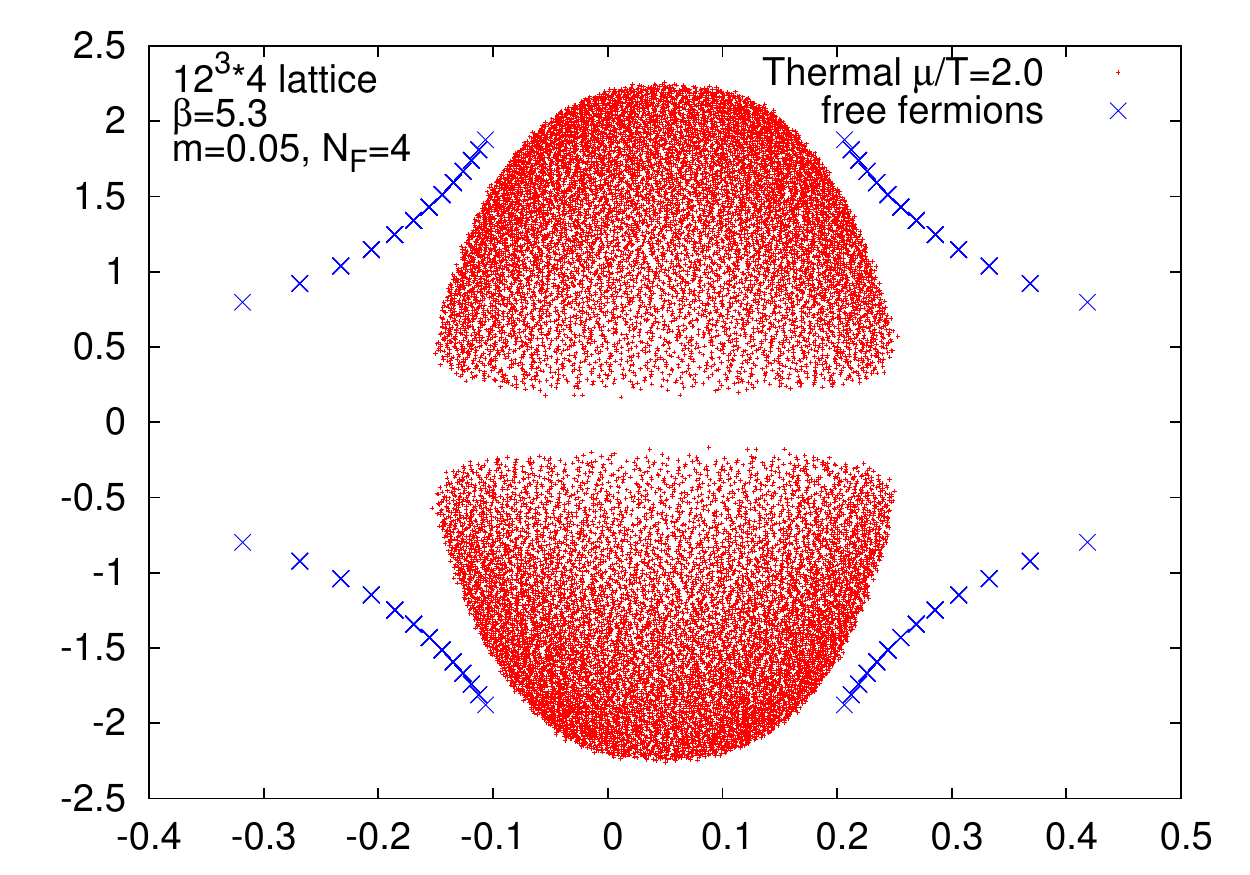}
\includegraphics[width=0.49\columnwidth]{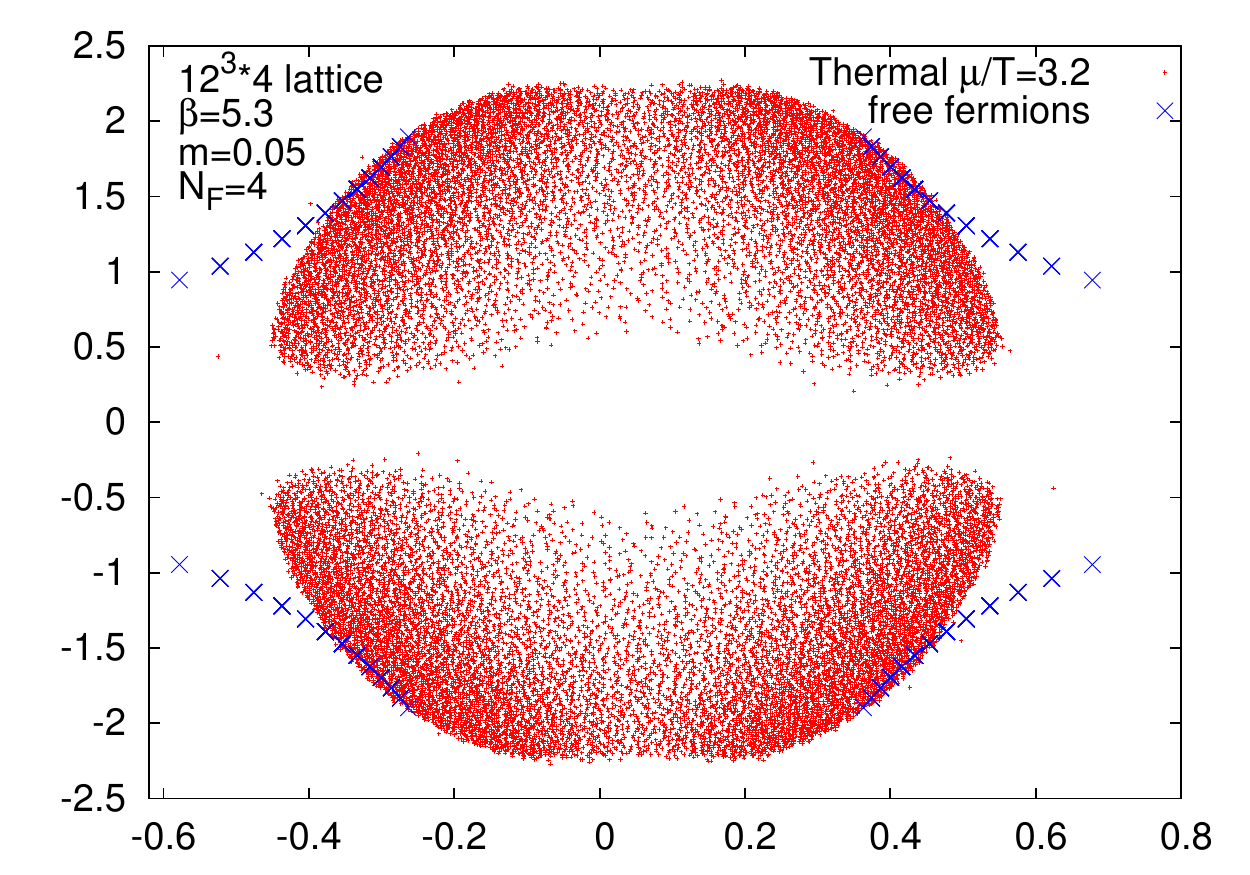}
\caption{The spectrum of the staggered Dirac operator in full QCD 
for various chemical potentials. The spectrum of the free 
staggered Dirac operator is also shown. 
}
\label{fullspectra}
\end{center}
\end{figure}

\begin{figure}[t]
\begin{center}
\includegraphics[width=0.69\columnwidth]{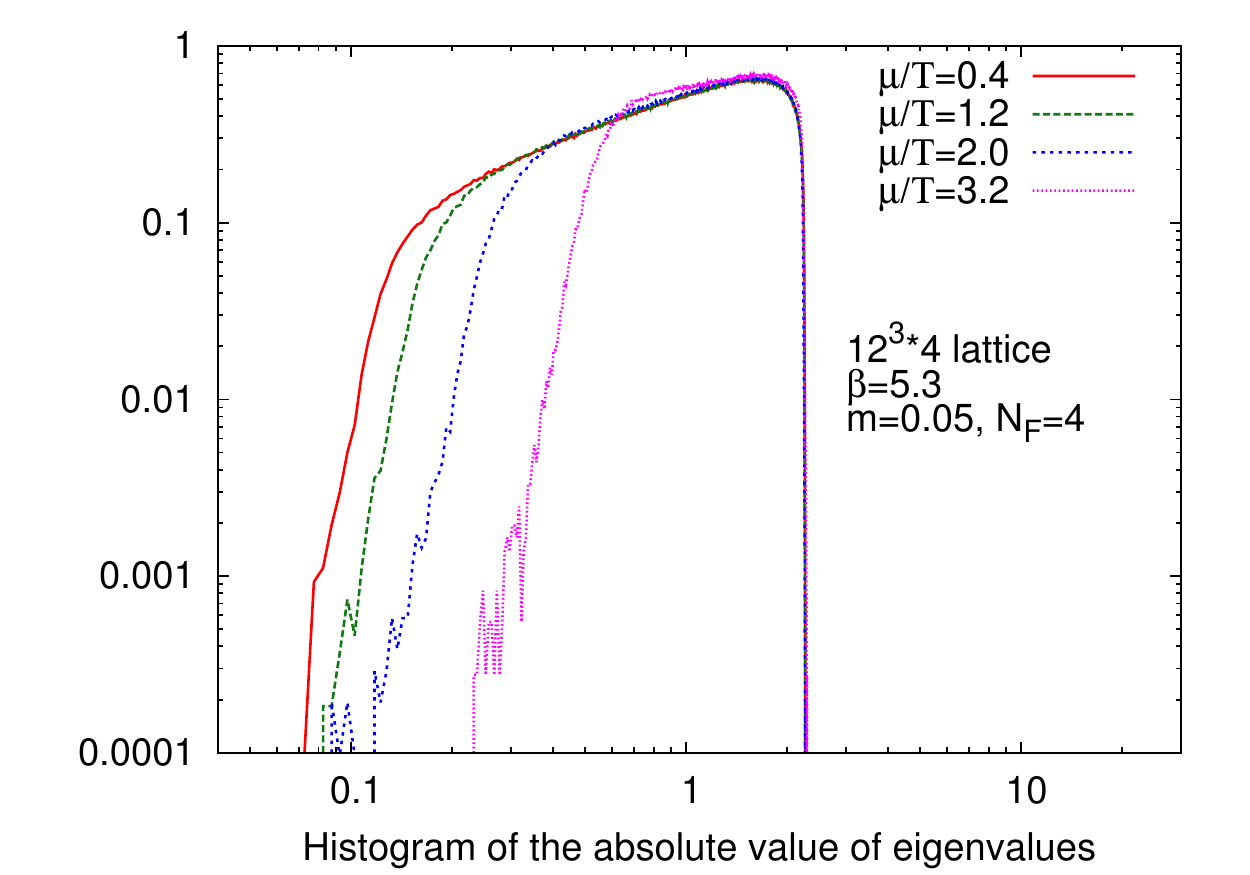}
\caption{The histogram of the absolute value of the eigenvalues 
of the staggered Dirac operator in full QCD. 
}
\label{fullrhist}
\end{center}
\end{figure}

\begin{figure}[t]
\begin{center}
\includegraphics[width=0.49\columnwidth]{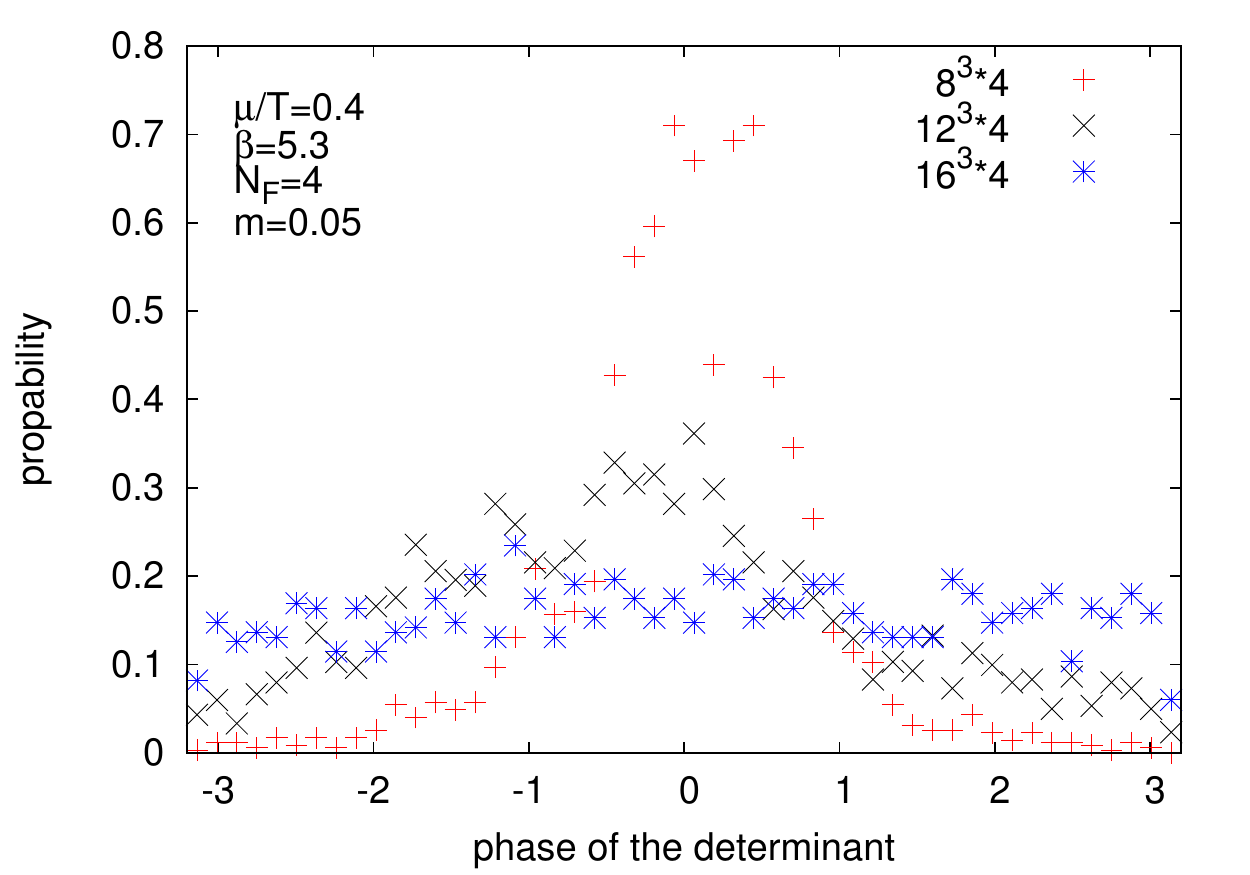}
\includegraphics[width=0.49\columnwidth]{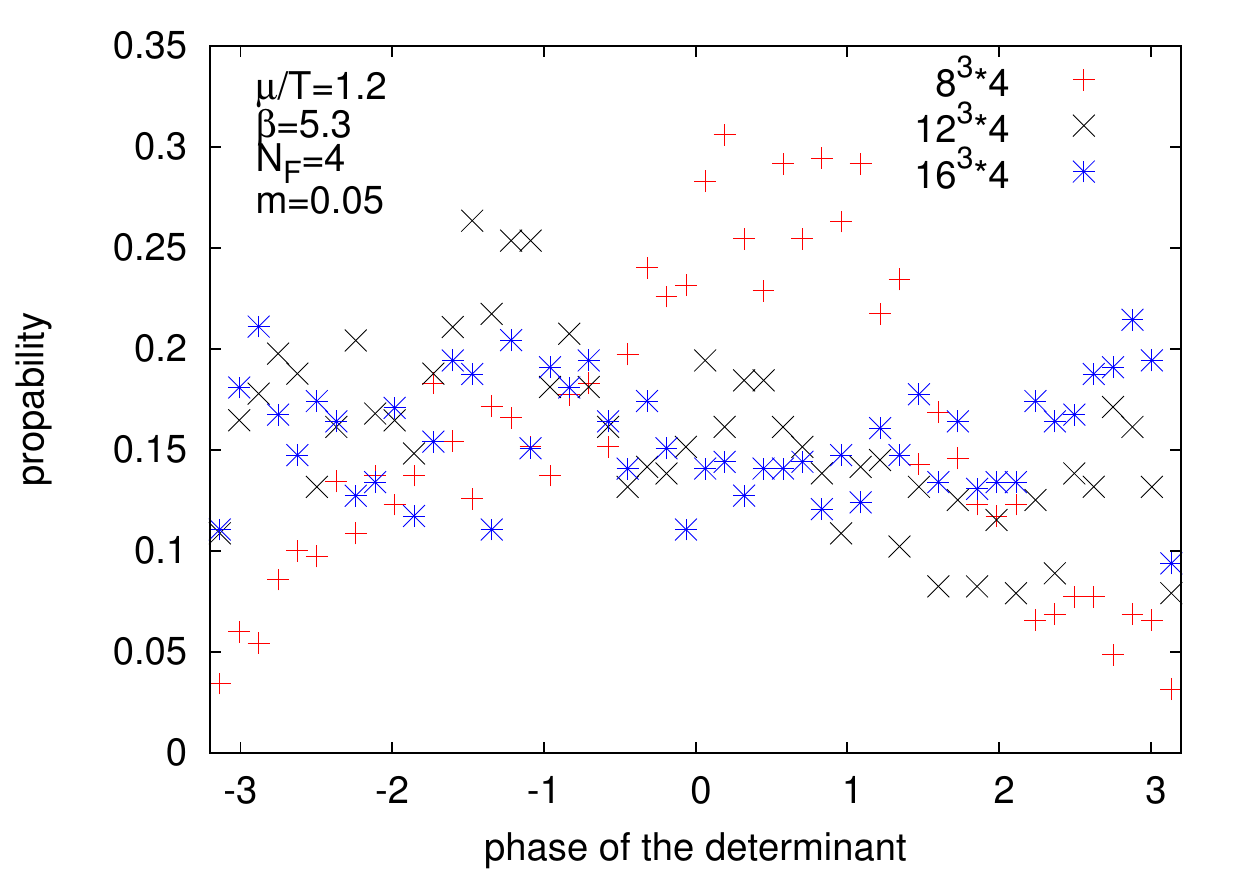}
\caption{The histogram of the phase of the determinant in full QCD using the staggered formulation.}
\label{fullphasehist}
\end{center}
\end{figure}

In full QCD the determinant can be written as the product of the eigenvalues of the 
Dirac matrix, and thus the fermionic force is written as 
$ K(U) = \sum_i D \lambda_i(U) / \lambda_i (U) $ with a sum for the eigenvalues 
$\lambda_i(U)$. Therefore a singularity in the drift term can be 
traced to a zero eigenvalue of the Dirac matrix. 
In Fig.~\ref{fullspectra} we show the 
spectrum of the complexified Dirac operator
on a $12^3 \times 4$ lattice at $\beta=5.3$ for various chemical 
potentials for typical configurations taken after sufficient Langevin time 
has passed to allow for thermalization.
One observes that the eigenvalues do not get close 
to zero. 
This behavior is also visible on the histogram of the absolute value of 
the determinants in Fig.~\ref{fullrhist}, where we taken 
an average of $O(100)$ configurations.
Note that this behavior does not mean that the phase of the full 
determinant remains small. In Fig.~\ref{fullphasehist} we show 
the histogram of the phase of the determinant for different spatial 
volumes at quark chemical potentials $ \mu/T = 0.4 \textrm{ and } 1.2 $.
Note that even at the smaller $\mu$ the phase distribution seems to be 
flat at $ N = 16$, making reweighting very hard. This behavior is expected 
as the sign average vanishes exponentially fast with the volume.

It is an open question whether at low temperature the zeroes of the measure 
affect the CLE results, as simulations are more expensive due to the larger 
lattices required there as the gauge cooling is ineffective on coarse lattices.

\section{Summary}

We have investigated the issue of the zeroes of the measure in the CLE solution
for non-zero density theories HDQCD and full QCD. The proof of correctness 
still holds as long as the singularity of the drift is outside of the 
distribution of the configurations. By numerical simulations we have shown 
that in HDQCD there is potentially a problem around the critical chemical 
potential of the onset transition but further study shows that 
the effect seems to be very small.
In full QCD at high temperatures we have studied the spectrum of the theory 
and shown that the eigenvalues of the Dirac matrix remain well separated 
from the origin for all chemical potentials.

\vskip 0.5cm
{\bf Acknowledgments }\ \  Support by the DFG (SE 2466/1-1) is gratefully acknowledged.
We also thank the Gauss Centre for Supercomputing for providing time on the JUQUEEN and SuperMUC.


\providecommand{\href}[2]{#2}\begingroup\raggedright\endgroup


\end{document}